\documentstyle[preprint,aps]{revtex}
\tightenlines
\begin{document}
\draft
\title{Truncation of large shell-model eigenproblems by
model space partitioning}
\author{F. Andreozzi and A. Porrino}
\address{Dipartimento di Scienze Fisiche, Universit\`a di Napoli Federico 
II,\\
and Istituto Nazionale di Fisica Nucleare. \\
Complesso Universitario di Monte S. Angelo, Via Cintia, 80126 Napoli, Italy}

\maketitle
\begin{abstract}
A method for solving the shell-model eigenproblem in a severely
truncated space, spanned by properly selected correlated states obtained
by partitioning the full configuration space, is proposed. The method  
describes in a practically exact way the low energy spectroscopic properties 
of nuclei, as exemplified in schematic models. The applicability of the 
method to heavy nuclei as well as
in contexts different from the nuclear shell model is stressed.
\end{abstract}

Complete shell-model calculations in the region of medium- and heavy-mass
nuclei become rapidly prohibitive because of the extremely large 
configuration space dimensions. Although the increasing power of the present 
day computers, togheter with a clever usage of the Lanczos algorithm\cite{Cau},
makes feasible exact calculations in configuration spaces of impressive
dimensions, the largest complete calculation in a full
major shell have been done, as to our knowledge, in the mass region around 
$^{56}$Ni. 
The extreme redundancy of the computed quantities is another 
problem: 
in the {\it mare magnum} of the output of these 
calculations, one is only 
interested in the comparison of a limited number of eigenvalues (usually
the lowest ones) with the observed energy levels and in the identification
of the most significant  components of the corresponding wave 
functions.
On the other hand, the truncation of the configuration space may not be
an easy task, since taking into account the effect of the excluded space
requires a corresponding renormalization of the interaction.

A main road to physically significant reductions of the
configuration space is the use of some kind of correlated basis;
typical examples are the multi-step shell model\cite{Lio},
the broken-pair approximation\cite{Bons} or the chain-calculation 
method\cite{CCM}.
The problem common to these  methods which make use of 
correlated bases is
the overcompleteness of the set of basis vectors used in the calculation.
Since this redundancy gives rise to spurious admixtures, it is necessary to
resort to special techniques to get rid of the spurious components, e.g. 
computing and
analyzing the overlap matrix of the basis, a task which is
lenghty and involved.

In this letter we propose a method which accounts in a practically exact way 
for all the configurations needed to the description of the lower energy states,
and is free of all the illnesses of the mentioned truncation schemes.
The idea consists in describing the system 
of interacting nucleons in terms 
of correlated subsystems defined in orthogonal subspaces.
Let us consider a system of $N$ valence particles in a
given model space, defined by a projection operator
$$ I = \sum_{l} \mid \alpha_l N \rangle \langle \alpha_l N \mid, \nonumber $$
where the $\alpha_l$ refer to all possible independent ways the $N$
particles can be distributed over the single-particle (s.p.) levels of 
the model space.
We now partition the s.p. states included in the model space  in two groups
whose corresponding configuration spaces we call $P_{1}(N_1)$ and 
$P_{2}(N_2)$, so that 
$$I = \sum_{N_1 N_2} \sum_{i j}\mid \beta^{(1)}_{i} N_1 \rangle \mid 
\beta^{(2)}_{j} N_2 
\rangle \;\; \langle \beta^{(2)}_{j} N_2 \mid \langle \beta^{(1)}_{i} 
N_1 \mid,$$
where the quantum numbers $\beta_i$ classify all
possible ways of distributing $N_1$ and $N_2$ particles, with $N_1+N_2 = N$,
on the s.p. levels of the 
two partitions, respectively.
We accordingly separate the Hamiltonian of the system as 
 $$H=H_1+H_2+H_{12},$$
\noindent 
where $H_1$ acts only on the  $P_1$ spaces, $H_2$ on the  $P_2$ spaces and
$H_{12}$ is the interaction term between $P_1$ and $P_2$.
Solving the eigenvalue problems for all the allowed values of $N_1$ 
and $N_2$
within each partition separately,
$$H_1 \; \mid \lambda_1 \; N_1 \rangle = E_{\lambda_1}(N_1) \; \mid 
\lambda_1 \; N_1 \rangle, \;\;\;\;
H_2 \; \mid \lambda_2 \; N_2 \rangle = E_{\lambda_2}(N_2) \; \mid 
\lambda_2 \; N_2 \rangle, \nonumber$$
\noindent
allows to write
\begin{equation}
I = \sum_{N_1 N_2} \sum_{\lambda_1 \lambda_2}\mid \lambda_{1} \; N_1 \rangle 
\mid \lambda_{2} \; N_2 \rangle \;\;
 \langle \lambda_{2} \; N_2 \mid \langle \lambda_{1} \; N_1 \mid.\label{base}
\end{equation}

Eq.(\ref{base}) is our main point. Since the wave function of 
the $N$-particle system 
is written as a direct product of eigenfunctions of correlated subsystems
defined in orthogonal spaces, none of the previously mentioned
redundancy problems arises, i.e. it combines the advantages of
using an orthonormal basis with those offered by a description in terms of
correlated subsystems. Shell-model basis is not well suited to further 
reductions of its dimensions essentially because the residual
interaction can be so strong that even configurations which are quite distant
in energy from each other may be equally important in the description
of a physical state. This is not the case when using correlated bases,
where a significant part of the effect of the interaction is already
included and even drastic truncations of the basis
can be meaningful. As a consequence, an energy criterion to truncate
expansion (\ref{base}) works well: only those basis states 
which are not too 
different in energy from the physical state one wants to describe are relevant.
It is worth noting that, as is explicitely shown in the following example, 
there are no particular
difficulties in computing the matrix elements of $H_{12}$ in the basis 
(\ref{base}), since they can be expressed in terms of quantities computed 
in each partition separately.

As a concrete example, let us consider a general two-body shell-model 
Hamiltonian for a system of identical particles:

\begin{equation}
H= \sum_{l} \epsilon_{l} \hat N_{l} 
+{1\over 4} \sum_{l_{1}l_{2}l_{3}l_{4} \atop J_0 M_0}
G_{J_0}^{\hphantom \dagger}
(l_{1}l_{2}l_{3}l_{4}) \; A_{J_0 M_0}^{\dagger}(l_{1}l_{2})
\; A_{J_0 M_0}^{\hphantom \dagger}(l_{3}l_{4}), 
\end{equation}
{\noindent where}
$\hat N_{l}= \sum_{m} a_{lm}^{\dagger} a_{lm}^{\hphantom \dagger}\>,$ and
$ A_{J_0M_0}^{\dagger}(l_{1}l_{2})= \sum_{m_{1}m_{2}} 
\langle l_{1}m_{1}l_{2}m_{2} \mid J_0M_0\rangle \; 
a_{l_{1}m_{1}}^{\dagger}a_{l_{2}m_{2}}^{\dagger}.$

\noindent{The partitions yield the two Hamiltonians}

\begin{equation}
H_{1} = \sum_{i} \epsilon_{i} \hat N_{i} 
+{1\over 4} \sum_{i_{1}i_{2}i_{3}i_{4} \atop J_0M_0}
G_{J_0}^{\hphantom \dagger}
(i_{1}i_{2}i_{3}i_{4}) \; A_{J_0M_0}^{\dagger}(i_{1}i_{2})
\; A_{J_0M_0}^{\hphantom \dagger}(i_{3}i_{4}),
\end{equation}

\begin{equation}
H_{2} = \sum_{j} \epsilon_{j} \hat N_{j} 
+{1\over 4} \sum_{j_{1}j_{2}j_{3}j_{4} \atop J_0M_0} 
G_{J_0}^{\hphantom \dagger}
(j_{1}j_{2}j_{3}j_{4}) \; A_{J_0M_0}^{\dagger}(j_{1}j_{2})
\; A_{J_0M_0}^{\hphantom \dagger}(j_{3}j_{4}),
\end{equation}

\noindent and the interaction term 

\begin{eqnarray}
H_{12} &=& {1\over 4} \sum_{i_{1}i_{2}j_{1}j_{2} \atop J_0M_0}
G_{J_0}^{\hphantom \dagger}
(i_{1}i_{2}j_{1}j_{2}) \;
A_{J_0M_0}^{\dagger}(i_{1}i_{2})
\; A_{J_0M_0}^{\hphantom \dagger}(j_{1}j_{2})
\nonumber \\
&+& {1\over 2} \sum_{i_{1}i_{2}j_{1}j_{2} \atop J_0 M_0} 
F_{J_0}^{\hphantom \dagger}(i_1i_2j_1j_2) 
\; B^{\dagger}_{J_0 M_0}(i_1i_2) \;
B^{\phantom\dagger}_{J_0 M_0}(j_1j_2)  \label{v12} \\
&+& {1\over 2} \sum_{i_{1}i_{2}i_{3}j_{1} \atop J_0} \hat{J}_0
F_{J_0}^{\hphantom \dagger}(j_1i_1i_2i_3)
\Biggl{[} \Big[ a^{\dagger}_{i_1} B^{\dagger}_{J_0}
(i_2i_3) \Big]^{j_1} \tilde{a}_{j_1} \Biggr{]}^{0}
\nonumber \\
&+& {1\over 2} \sum_{j_{1}j_{2}j_{3}i_{1} \atop J_0} \hat{J}_0
F_{J_0}^{\hphantom \dagger}(i_1j_1j_2j_3)
\Biggl{[} \Big[ a^{\dagger}_{j_1} B^{\dagger}_{J_0}
(j_2j_3) \Big]^{i_1} \tilde{a}_{i_1} \Biggr{]}^{0} + h.c.,
\nonumber
\end{eqnarray}

\noindent where $\tilde{a}_{lm} = (-1)^{l+m}a_{l-m}$ and 
$\hat{J}_0 = (2J_0+1)^{1/2}$. 
The $ F $ matrix elements  are defined as

\begin{equation}
F_{J_0}^{\hphantom \dagger}(l_{a}l_{b}l_{c}l_{d}) 
= - \sum_{J_{1}} \; (2 J_{1} + 1) \;
G_{J_{1}}^{\hphantom \dagger}(l_{a}l_{d}l_{c}l_{b})
\left\{ \begin{array}{ccc} l_{a} & l_{b} & J_0 \\ l_{c} & l_{d} & J_{1}
\end{array} \right\},
\end{equation}

\noindent{and} $B_{J_0M_0}^{\dagger}(l_1 l_2)$ are the particle-hole
operators:

\begin{equation}
 B_{J_0M_0}^{\dagger}(l_{1}l_{2})= \sum_{m_{1}m_{2}} 
\langle l_{1}m_{1}l_{2}m_{2} \mid J_0M_0\rangle \; 
a_{l_{1}m_{1}}^{\dagger} \tilde{a}_{l_{2}m_{2}}^{\phantom{\dagger}}.
\end{equation}

Let $\mid N_1 \gamma_1 J_1 M_1\rangle$ be the eigenstates of $H_1$ for
a system of $N_{1}$ particles distributed over the s.p.
levels of the partition 1 of the model space,
with corresponding eigenvalues $E_{\gamma_1 J_1}(N_1)$, and
$\mid N_2 \gamma_2 J_2 M_2 \rangle$ the eigenstates of $H_2$ for
a system of $N_{2}$ particles distributed over the 
s.p. levels of the partition 2,
with corresponding eigenvalues $E_{\gamma_2J_2}(N_2)$.

A complete  basis for the configuration space of $N$ identical particles, 
distributed over all the s.p. levels of the model space,  
with total angular momentum $JM$,
can be therefore written as

\begin{equation}
\mid N_1,\gamma_1,J_1,N_2,\gamma_2,J_2,N JM \rangle =
\sum_{M_1M_2} \langle J_1 M_1 J_2 M_2 \mid JM \rangle \;
\mid N_1 \gamma_1 J_1 M_1\rangle
\mid N_2 \gamma_2 J_2 M_2 \rangle.\label{basesm}
\end{equation}

\noindent where $N_1$ and $N_2$ run over all the possible values such that 
$ N_1 +N_2 = N$.

To exemplify the quality of the results obtainable
when an energy truncation criterion is used, we consider for simplicity
a pairing Hamiltonian 

\begin{equation}
H= \sum_{l} \epsilon_{l} \hat N_{l} 
+{1\over 4} \sum_{l l'} 
G_{0}^{\hphantom \dagger}
(l,l') \; A_{0}^{\dagger}(l)
\; A_{0}^{\hphantom \dagger}(l'), 
\end{equation}

\noindent{where we use the shorthand}
$G_{0}^{\hphantom \dagger}(l,l') \equiv
G_{0}^{\hphantom \dagger}(lll'l')$ and
$A_{0}^{\dagger}(l) \equiv  A_{00}^{\dagger}(ll),$
and restrict ourselves to a seniority zero approximation for the
configuration space.
The interaction term (\ref{v12}) becomes
\begin{equation}
H_{12} = {1\over 4} \sum_{i j} 
G_{0}^{\hphantom \dagger}
(i,j) \; A_{0}^{\dagger}(i)\; A_{0}^{\hphantom \dagger}(j) + h.c. .
\end{equation}

\noindent The basis for an $N$-particle system is now 
$\mid N_1 \gamma_1, N_2 \gamma_2; N \rangle \; = \;
 \mid N_1 \gamma_1 \rangle \mid N_2 \gamma_2 \rangle.$
The matrix elements of $H$ between those basis states
are:

\begin{eqnarray}
\langle N'_1 \gamma'_1,N'_2 \gamma'_2; N \mid &H&\;
\mid N_1 \gamma_1,N_2 \gamma_2; N \rangle = 
\Big[ E_{\gamma_1}(N_1) + E_{\gamma_2}(N_2) \Big]
\delta_{N_1' N_1^{\phantom{i}}}
\delta_{N_2' N_2^{\phantom{i}}}
\delta_{\gamma_1' \gamma_1^{\phantom{i}}}
\delta_{\gamma_2' \gamma_2^{\phantom{i}}} \\ 
&+& {1\over 4} \sum_{ij}
G_{0}^{\hphantom \dagger}(ij) \;\Biggl{[} 
X_{i \gamma_1' \gamma_1^{\phantom{i}}}(N'_1) \;
X_{j \gamma_2^{\phantom{i}} \gamma_2'}(N_2) \;
\delta_{N_1' (N_{1}^{\phantom{i}}+2)} \; 
\delta_{N_2^{\phantom{i}}(N_{2}'+2)} \nonumber \\
&\;& \;\;\;\;\;\;\;\;\;\;\;\;\;\;\;\;\; + \; 
X_{i \gamma_1^{\phantom{i}}\gamma_1'}(N_1) \;
X_{j \gamma_2' \gamma_2^{\phantom{i}}}(N_2') \;
\delta_{N_1^{\phantom{i}}
 (N_{1}'+2)} \; 
\delta_{N'_2 (N_{2}^{\phantom{i}}+2)} \Biggr{]}, \nonumber
\end{eqnarray}
\noindent where $X_{k \gamma^{\phantom{i}}_l \gamma'_l}(N_l)
= \langle N_l \; \gamma_l^{\phantom{i}} \mid \mid A_{0}^{\dagger}(k) \mid 
\mid (N_{l}-2) \;  \gamma_l' \rangle$ are two-particle transfer amplitudes.

\noindent As already pointed out, matrix elements of $H_{12}$ are quite
simple in structure and can be evaluated using quantities defined in
each subspace separately. It is worth noting that this is not
peculiar of a pairing interaction:
as can be easily seen from the structure of Eq.(\ref{v12}),
matrix elements of $H_{12}$ in the basis (\ref{basesm})
can be written in terms of the the two-particles transfer amplitudes
$X_{l_1 l_2 J_0 \gamma_l J_l \gamma'_l J'_l}(N_l) =
\langle N_l, \gamma_l, J_l \mid\mid A^{\dagger}_{J_0}(l_1l_2) \mid\mid
N'_l, \gamma'_l, J'_l \rangle $,
of the particle-hole matrix elements
$\langle N_l, \gamma_l, J_l \mid\mid B^{\dagger}_{J_0}(l_1l_2) \mid\mid
N_l, \gamma'_l, J'_l \rangle$
and of the one-particle transfer amplitudes
$\langle N_l,\gamma_l, J_l\mid \mid a^{\dagger}_{l_1}
\mid \mid N'_l,\gamma'_l,J'_l \rangle$, computed in each partition separately.

To reduce the dimension of the eigenvalue problem we retain only 
the basis vectors 
$\mid N_1 \gamma_1, N_2 \gamma_2; N \rangle$
corresponding to values of 
$\big{[}E_{\gamma_1}(N_1) + E_{\gamma_2}(N_2)\big{]}$ up to a truncation
value $E^{*}$:
progressively increasing $E^*$, one can see how the energies in the full space
approximate the exact ones. 
We have chosen a
model space of ten s.p. levels divided in two partitions of five s.p. levels.
In Table I are reported the first three energies obtained with various 
truncations of the basis with $N = 30$ nucleons. The ten s.p. levels are 
$0f_{7/2} , 0f_{5/2} , 1p_{3/2} , 1p_{1/2} , 0g_{9/2}$, 
which define the partition 1, and
$0g_{7/2} , 1d_{5/2} , 1d_{3/2} , 2s_{1/2} , 0h_{11/2}$, which define
the partition 2. The corresponding s.p. energies are, in MeV, 0.0, 3.0, 3.5,
4.0, 5.0, 13.0, 13.5, 15.0, 15.5, 16.0 and we use a constant pairing strength
$ G = 0.16$ MeV, where $G_0(l l') = - \hat{l} \hat{l}' G$.
We obtain practically exact results already
diagonalizing matrices of order 150 while the dimension of
the full space basis is $59,702$.

Just to show that the existence of a gap in the s.p. energies is
not essential to the quality of the approximation, we have also
considered the simple, although not at all trivial, example of a pairing
problem in a model space of 32 equispaced doubly-degenerate levels. Exact 
solutions of this problem in the case of 32 particles have been found in 
Ref.\cite{Rich} for the energies of the first three states. We arbitrarily 
choose as partition 1 the first 16 levels and as partition 2 the remaining 
ones. We have generated almost exact solutions 
for each value of $N_1$ and $N_2$ within each subspace, 
with the corresponding two-particle transfer amplitudes, 
using the method of Refs. \cite{CCM} \cite{def}.
The results, reported in Table II, 
show an impressive
convergence towards the exact values: the best computed  energies 
differ by less than 0.1\% from the exact ones, while the corresponding
excitation energies, 3.060 and 4.900, are practically exact. 
It is worth noting that we use a number of states 
which is an exceedingly small fraction of the full basis dimension 
$6.01 \times 10^8$.
These results are
significantly better than those of Ref.\cite{Dud}, where a careful analysis of 
truncations of the configuration space is made; this fact not
only confirms the effectiveness of the present method but
also suggests that it can be a viable technique to obtain almost exact 
solutions  for other physically interesting systems like those encountered
in the modelization of ultra-small superconducting grains\cite{faz}\cite{Braun}.

We have considered in the preceding discussion only two partitions, just to 
be definite. One can have, however, situations 
where a basis generated by multiple partitioning the configuration space 
would be required. 
Althoug we do not treat these cases explicitely in this paper, it is worth
noting that the generalization of the method presented here to systems
where multiple partitioning is needed is quite straightforward.
The method can be naturally applied to systems of 
neutrons and protons, which can be conveniently treated working out the
eigenproblems for neutrons and protons separately by further partitioning of
their model spaces, and then diagonalizing the n-p interaction in the 
product basis; it also constitutes a natural framework to treat
excitations outside a major shell,
e.g. in the study of ground state correlations.

{\bf Acknowledgements} We thank N. Lo Iudice for fruitful discussions and
comments.
This work was supported in part by the Italian Ministero dell' Universit\`a
e della Ricerca Scientifica e Tecnologica (MURST).

\begin{table}
\caption{Absolute energies, in MeV, of the first three $v = 0$ states, 
$N = 30$, for the model space of ten s.p. levels. 
The first row gives the number of states to which the full basis is truncated.}
\begin{tabular}{ddddddd}
 &25 &60 &150 &230 &430 & 810\\
\tableline
$E_0$ &90.256& 90.244 & 90.241 &90.241& 90.241 & 90.241 \\
$E_1$ &104.286&  104.065 &103.990 &103.972& 103.962 & 103.960\\
$E_2$ & 106.474& 106.058 & 106.010 &105.996 & 105.991 &105.989\\
\end{tabular}
\label{table1}
\end{table}

\begin{table}
\caption{Absolute energies, in MeV, of the first three states
 for the model space of 32 twofold-degenerate equispaced levels,
$N = 32$ particles, level spacing $1$ MeV, pairing constant $G = 0.345$
MeV. The first row gives the number of states to  which the full basis is
truncated.}
\begin{tabular}{ddddddd}
 & 1057 & 2078 & 3120 & 3454 & 3999 & exact\\
\tableline
$E_0$ & 263.587 & 263.453 & 263.390 & 263.375 & 263.356 & 263.171 \\
$E_1$ & 266.599 & 266.498 & 266.443 & 266.434 & 266.416 & 266.278 \\
$E_2$ & 268.496 & 268.352 & 268.292 & 268.271 & 268.256 & 268.071 \\
\end{tabular}
\label{table2}
\end{table}

\end{document}